\documentclass{article}
\usepackage{spconf,amsmath,graphicx}

\usepackage{url}
\usepackage{amsmath}
\usepackage{amsfonts}
\usepackage{amssymb}
\usepackage{amsthm}

\usepackage{cite}
\usepackage{multirow}
\usepackage[bottom]{footmisc}
\usepackage{subfigure}
\usepackage{graphicx}
\usepackage{caption}
\usepackage[labelformat=simple]{subfig}
\usepackage{enumitem}

 \usepackage{amsthm}
 \theoremstyle{plain}

\theoremstyle{definition}
\newtheorem{definition}{Definition}

\DeclareMathOperator*{\minimize}{\mathrm{minimize}}

\DeclareMathOperator{\subjto}{\mathrm{subject \, to}}

\newcommand{\tran}{^{\scriptscriptstyle{\sf T}}} 

\newcommand{\Graph}{\mathcal{G}}

\newcommand{\Adj}{{\mathbf{W}}} 
\newcommand{\SLoop}{\mathbf{V}} 
\newcommand{\Degree}{\mathbf{D}} 

\newcommand{\lap}{{\mathbf{L}}}

\newcommand{\eigM}{{\mathbf{\Lambda}}}

\newcommand{\CS}{\mathbf{S}} 

\newcommand{\gbt}{{\mathbf{U}}} 

\usepackage{enumitem}

\setitemize[0]{leftmargin=15pt,itemindent=0pt}

\ninept
\title{Parametric Graph-based Separable Transforms for Video Coding}
\name{Hilmi E. Egilmez$^\ast$, Oguzhan Teke$^\dagger$, Amir Said$^\ast$, Vadim Seregin$^\ast$ and Marta Karczewicz$^\ast$ \thanks{This work was supported by Qualcomm Technologies, Inc. O. Teke was with Qualcomm Technologies, Inc., while most of this work has been done. He is currently with the Department of Electrical Engineering, California Institute of Technology, Pasadena, CA 91125 USA.}}
\address{$^\ast$Qualcomm Technologies, Inc., San Diego, CA 92121, USA.\\
         $^\dagger$California Institute of Technology, Pasadena, CA 91125 USA.}
\begin{document}
%
\maketitle
\begin{abstract}
In many video coding systems, separable transforms (such as two-dimensional DCT-2) have been used to code block residual signals obtained after prediction. This paper proposes a parametric approach to build graph-based separable transforms (GBSTs) for video coding. Specifically, a GBST is derived from a pair of line graphs, whose weights are determined based on two non-negative parameters. As certain choices of those parameters correspond to the discrete sine and cosine transform types used in recent video coding standards (including DCT-2, DST-7 and DCT-8), this paper further optimizes these graph parameters to better capture residual block statistics and improve video coding efficiency. The proposed GBSTs are tested on the Versatile Video Coding (VVC) reference software, and the experimental results show that about 0.4\% average coding gain is achieved over the existing set of separable transforms constructed based on DCT-2, DST-7 and DCT-8 in VVC. 
\end{abstract}
\urlstyle{same}

\section{Introduction}
\label{sec:intro}
In most video compression systems predating the \emph{High-Efficiency Video Coding} (HEVC) standard \cite{Sullivan:12:hevc}, only the type-2 discrete cosine transform (DCT-2) is used to transform rows and columns of a residual block signal in a separable manner. The main problem of using DCT-2 as the only transform option is the implicit assumption that all residual blocks share the same statistical properties. However, in practice, the residual blocks can have diverse statistical characteristics depending on video content, prediction mode and block size. It has been shown that the separable type-7 discrete sine transform (DST-7) can provide considerable coding gains over DCT-2 for small blocks \cite{Han:12:hybrid}, and DST-7 is adopted in HEVC for coding $4\!\times\!4$ intra predicted blocks \cite{Sullivan:12:hevc}. In order to further increase the diversity in transform selection with a significant coding benefit, 
the AV1 standard \cite{av1_standard} enables combinations of DST-7 and DCT-8 for block sizes up to $16\!\times\!16$. Moreover, the \emph{Versatile Video Coding} (VVC) standard, currently under development stage \cite{Bross:19:vvc7}, employs \emph{multiple transform selection} (MTS) among five candidates derived based on DCT-2, DCT-8 and DST-7, where larger block transforms up to size $32\!\times\!32$ are supported \cite{Said:18:CE6_report}. This paper proposes a learning method to design parametric \emph{graph-based separable transforms (GBSTs)} as generalized extensions of DST-7 and DCT-8 for the MTS design in VVC \cite{Bross:19:vvc7}.

{Graph-based transforms (GBTs)} \cite{egilmez:15:gbt_inter,pavez:2015:gtt_paper,egilmez:16:gbst,egilmez:2019:graph_video_arxiv} have been introduced to improve coding efficiency by providing a better adaptation for diverse block statistics. Particularly in \cite{egilmez:16:gbst,egilmez:2019:graph_video_arxiv}, {separable GBTs (GBSTs)} are optimized by learning a pair of line graphs from data \cite{egilmez:2017:gl_from_data_jstsp} to model underlying row and column-wise statistics of blocks residual signals, where the associated graph Laplacian matrices are used to derive GBSTs. 
In \cite{egilmez:2019:graph_video_arxiv}, the authors further show that GBTs can outperform separable Karhunen-Loeve transforms (KLTs) derived from sample covariance matrices in terms of coding efficiency, 
since the graph learning approach provides a better model estimation (i.e., better \emph{generalization} and \emph{variance-bias trade-off} \cite{vapnik:1999:slt,loxburg:2011:slt}) than using sample covariances for deriving KLTs. 
One of the key conclusions in \cite{egilmez:2019:graph_video_arxiv} is that learning models with fewer parameters (which alleviates the \emph{overfitting} problem) often results in a more general and robust model estimation, so that graph learning techniques can be employed to design better transforms for video coding. 
Based on this result, this paper introduces a parametric approach to learn line graphs, whose weights are determined from two non-negative parameters. Contributions of the present paper can be summarized as follows:
\begin{itemize}[noitemsep,nolistsep]
    \item A constrained optimization framework with a new training method is proposed to learn parameters of line graphs used to derive GBSTs.
    \item Practical GBSTs at 8-bit integer precision are constructed and tested by replacing the transform matrices of MTS in VVC. A comprehensive empirical analysis of graph parameter choices (in terms of coding efficiency) is also presented.
    \item The proposed GBSTs do not introduce an additional hardware/software complexity to the current VVC design, since common implementations of DST-7 and DCT-8 are based on matrix multiplications.
\end{itemize}
As compared to the existing GBST construction methods allowing arbitrary graph weights in \cite{egilmez:2019:graph_video_arxiv}, the proposed parametric approach is a more constrained variant leading to simpler models defined by fewer parameters. 
Besides, the empirical results in \cite{egilmez:16:gbst,egilmez:2019:graph_video_arxiv} are provided in a simple simulation setting with their theoretical justifications. 
However, in the present paper, proposed parametric GBSTs are evaluated in a more practical setting, where they are tested on the VVC reference software by benchmarking against the existing MTS scheme in VVC with DST-7 and DCT-8 \cite{Said:18:CE6_report}.  

The rest of the paper is organized as follows. Section \ref{sec:prelim} gives some preliminary information and definitions. 
In Section \ref{sec:approach}, the problem formulation and proposed parametric approach are presented. The experimental results are discussed in Section \ref{sec:results}, and Section \ref{sec:concl} draws some conclusions.

\section{Preliminaries}
\label{sec:prelim}
In this paper, the graphs of interest are algebraically represented by generalized graph Laplacian (GGL) matrices.
\begin{definition}[GGL]
Given a weighted graph $\Graph(\Adj,\SLoop)$, the generalized graph Laplacian is defined as
\begin{equation}
\lap = \Degree - \Adj + \SLoop
\label{eqn:gen_lap_form}
\end{equation}  
where $\Adj$ is the adjacency matrix consisting of non-negative edge weights, $\Degree$ is the diagonal degree matrix, and $\SLoop$ is the diagonal matrix denoting weighted self-loops (i.e., vertex-weights). Note that the $\Degree - \Adj$ term is the combinatorial Laplacian matrix \cite{Chung:1997:SGT}, and GGL is obtained by adding the vertex weights in $\SLoop$.  
\end{definition}

\noindent A GBST is defined by a pair of GBTs derived from the GGLs associated with two weighted line graphs.  We formally define GBT and GBST as follows.
\begin{definition} [GBT] GBT of a graph $\Graph(\Adj,\SLoop)$ is obtained by eigen-decomposition of the generalized graph Laplacian,
\begin{equation}
\label{eqn:GBT}
\lap = \Degree - \Adj + \SLoop = \gbt \eigM \gbt\tran,
\end{equation} 
where columns of orthogonal matrix ${\mathbf{U}}$ are the basis vectors of the GBT, and ${\mathbf{\Lambda}}$ is the diagonal eigenvalue matrix.
\end{definition}

\begin{definition} [GBST] Let $\gbt_{\textnormal{row}}$ and $\gbt_{\textnormal{col}}$ be $N \times N$ GBTs associated with two line graphs with $N$ vertices, then the GBST of an $N \! \times \! N$ matrix $\mathbf{X}$ is  
\begin{equation}
\widehat{\mathbf{X}} = \gbt_{\textnormal{col}}\tran \; \mathbf{X} \; \gbt_{\textnormal{row}}, 
\label{eqn:gbst_video}
\end{equation}
where $\gbt_{\textnormal{row}}$ and $\gbt_{\textnormal{col}}$ 
are the transforms applied rows and columns of the block signal $\mathbf{X}$, respectively.
\end{definition}

\begin{figure}[!t]
\centering
\subfigure[Graph weights corresponding to GGLs in $\mathcal{L}_1$]{\includegraphics[width=0.35\textwidth]{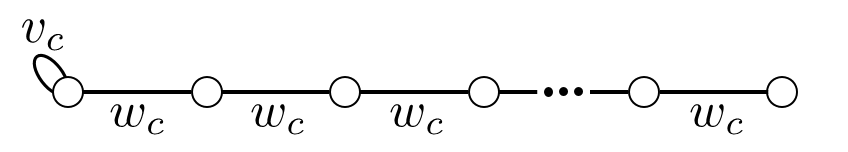}}
\subfigure[Graph weights corresponding to GGLs in $\mathcal{L}_2$]{\includegraphics[width=0.35\textwidth]{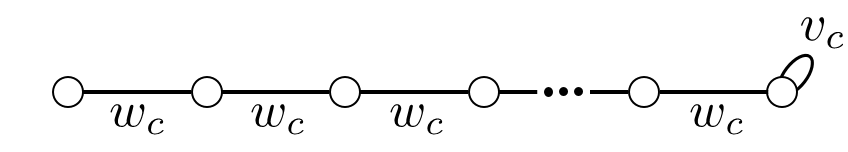}}
\caption{Graphs of interest with constant edge weights ($w_c$) and a single self-loop weight ($v_c$) at (a) the first vertex and the (b) last vertex.} 
\label{fig:line_graphs}
\end{figure}

\section{Proposed Parametric Framework for GBST Construction}
\label{sec:approach} 
In this section, we first define the form of parametric GGLs considered in this paper and discuss their relations with different types of DCT/DST. Then, the problem formulation and proposed solution are presented. 
\subsection{Target GGLs representing line graphs}
\label{sec:target_ggls}
In this paper, the following two sets of GGL matrices representing the line graphs are considered to derive GBSTs:
\begin{equation}
\mathcal{L}_1 \!=\! \left\lbrace \!
\left. \!
\begin{bmatrix}
w_c\!+\!v_c & -w_c &  &   & 0  \\
-w_c & 2w_c & -w_c &   \\
  & \ddots & \ddots & \ddots  \\
   & & -w_c &2 w_c& -w_c  \\
0 &  & & -w_c & w_c  
\end{bmatrix} \! \right|  
\begin{aligned} 
w_c \geq 0 \\
v_c \geq 0
\end{aligned} \right\rbrace
\label{eqn:set_L1}
\end{equation}
and
\begin{equation}
\mathcal{L}_2 \!=\! \left\lbrace \!
\left. \!
\begin{bmatrix}
w_c & -w_c &  &   & 0  \\
-w_c & 2w_c & -w_c &   \\
  & \ddots & \ddots & \ddots  \\
   & & -w_c &2 w_c& -w_c  \\
0 &  & & -w_c & w_c\!+\!v_c  
\end{bmatrix} \! \right|  
\begin{aligned} 
w_c \geq 0 \\
v_c \geq 0
\end{aligned} \right\rbrace
\label{eqn:set_L2}
\end{equation}
where $w_c$ denotes the constant edge weights of the line graphs, and $v_c$ is the vertex weight (i.e., self loop weight). As illustrated in Fig.~\ref{fig:line_graphs}, the only structural difference in GGL sets is the location of weighted vertex (i.e., self-loop).

\subsection{DCTs and DSTs as GBTs derived from line graphs}
\label{sec:dct_dst_interpret}
Some types of DCTs and DSTs, including DCT-2, DCT-8 and DST-7, are in fact GBTs derived from certain forms of GGLs. 
Based on the results in \cite{Strang:1999:DCT,Moura_Pueschel:08:algebraic_signal_proc_space,egilmez:16:gbst,gene:15:GGL}, the parameters $v_c $ and $w_c$ can be selected such that the specified sets of GGLs in (\ref{eqn:set_L1}) and (\ref{eqn:set_L2}) correspond to different types of DCTs and DSTs. For example, the relation between different types of DCTs and graph Laplacians is originally discussed in \cite{Strang:1999:DCT} where DCT-2 is shown to be equal to the GBT uniquely obtained from combinatorial graph Laplacians of the following form
\begin{equation}
\lap_c \!=\! \begin{bmatrix}
w_c & -w_c &  &   & 0  \\
-w_c & 2w_c & -w_c &   \\
  & \ddots & \ddots & \ddots  \\
   & & -w_c &2 w_c& -w_c  \\
0 &  & & -w_c & w_c 
\end{bmatrix} \text{ for $w_c>0$},
\label{eqn:CGL_form_DCT2}
\end{equation}
which represents uniformly weighted line graphs with no self-loops (i.e., all edge weights are equal to a positive constant and vertex weights to zero). Then, the form in (\ref{eqn:CGL_form_DCT2}) can be obtained by setting $v_c = 0$ with $w_c > 0$ for the target GGL sets $\mathcal{L}_1$ and $\mathcal{L}_2$. By combining this observation with the analysis in \cite{Moura_Pueschel:08:algebraic_signal_proc_space,egilmez:16:gbst,gene:15:GGL}, different choices of $v_c $ and $w_c$ for the GGL sets $\mathcal{L}_1$ and $\mathcal{L}_2$ result in the following five DCT/DST types:
\begin{itemize}
\item DCT-2 is derived by setting $v_c=0$ in $\mathcal{L}_1$ or $\mathcal{L}_2$. 
\item DST-7 is derived by setting $v_c=w_c$ in $\mathcal{L}_1$.
\item DCT-8 is derived by setting $v_c=w_c$ in $\mathcal{L}_2$.
\item DST-4 is derived by setting $v_c=2w_c$ in $\mathcal{L}_1$.
\item DCT-4 is derived by setting $v_c=2w_c$ in $\mathcal{L}_2$.
\end{itemize}
Moreover, it is important to note that DST-7 and DCT-8 are closely related transforms, where the basis vectors of one are in fact the reversed (flipped) versions of the other \cite{Said:19:taf}, as the only difference in their associated graphs is the location of the self-loop (see Fig.~\ref{fig:line_graphs})\footnote{Mathematically, the graphs shown in Fig.~\ref{fig:line_graphs} are isomorphic to each other.}. The DST-4 and DCT-4 also share this property. 

\subsection{Problem Formulation}
In order to learn graphs from data, the following parametric GGL estimation problem is formulated:
\begin{equation}
 \begin{aligned}
& \minimize_{w_c \geq 0, v_c \geq 0}
& & 
 \mathrm{Tr} \left( \lap(w_c,v_c) \CS \right) -\mathrm{logdet}  ( \lap(w_c,v_c) )  \\
& \subjto
& &   \lap(w_c,v_c) \in \mathcal{L}_t \\ 
\end{aligned}
\label{eqn:prob_formulation}
\end{equation}
where $\CS$ denotes the sample covariance matrix obtained from data, and $\lap(w_c,v_c)$ is the GGL variable defined by parameters $v_c$ and $w_c$. The operators $\mathrm{Tr}(\cdot)$ and $\mathrm{logdet}(\cdot)$ correspond to the matrix trace and determinant of logarithm, respectively. 
Since the minimization criterion in (\ref{eqn:prob_formulation}) is based on negative log-likelihood of a Gaussian-Markov random field (GMRF), whose precision matrix is a GGL \cite{egilmez:2017:gl_from_data_jstsp}, it is a constrained maximum-likelihood (ML) problem for estimating a GGL over the constraint set $\mathcal{L}_t$ defined by the parameters $v_c$ and $w_c$. 

This paper focuses on solving the instances of (\ref{eqn:prob_formulation}) where the constraint set $\mathcal{L}_t$ can be either $\mathcal{L}_1$ or $\mathcal{L}_2$, defined in Section \ref{sec:target_ggls}. Since (\ref{eqn:prob_formulation}) is a convex optimization problem \cite{boyd:2005:cvx_book} and the dimension of the problem is small, the CVX software \cite{cvx_software} is used to find the optimal solutions for $w_c$ and $v_c$. As a more efficient alternative to CVX, the \emph{GGL estimation algorithm} proposed in \cite{egilmez:2017:gl_from_data_jstsp,GLL_package:v1.0} can be modified to accommodate the set constraints in $\mathcal{L}_1$ and $\mathcal{L}_2$ by introducing projected descent steps. Such efficient alternatives are out of the scope of this paper and considered as part of the future work. 

\begin{figure*}[!t]
\centering
\includegraphics[trim={5cm 1cm 0cm 0cm},clip,width=\textwidth]{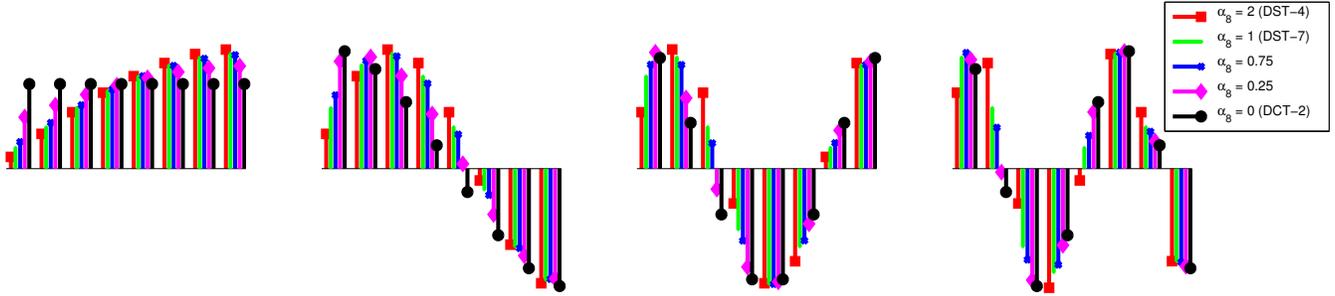}
\caption{Illustration of the four lowest frequency basis vectors for 8-point transforms derived from graphs with $\alpha_8=2$, $\alpha_8=1$, $\alpha_8=0.75$, $\alpha_8=0.25$ and $\alpha_8 = 0$. The transforms obtained based on $\alpha_8=2$, $\alpha_8=1$ and $\alpha_8=0$ are equal to 8-point DST-4, DST-7 and DCT-2, respectively.} 
\label{fig:basis}
\end{figure*}

\subsection{Proposed Method}
\label{sec:proposed_approach}
To design GBSTs, we propose a two-step procedure to optimize a GGL matrices used to derive transforms. In the first step, the problem in (\ref{eqn:prob_formulation}) is solved to find the optimal parameters ($w_c^\ast$ and $v_c^\ast$) that best capture the input data statistics (sample covariance $\CS$) in a maximum-likelihood sense. The residual data is collected by using the VVC reference software (version 2), VTM, where the row and column statistics ($\CS$) are obtained from intra coded residual blocks\footnote{The video sequences used in HEVC standardization (different from the test sequences in VVC) are used for generating the training dataset.}. After solving (\ref{eqn:prob_formulation}), the second step refines the optimized GGL matrix, $\lap(w_c^\ast,v_c^\ast$) by normalizing entries of the GGL and then rounding the normalized vertex weight to a value of admissible precision\footnote{The refinement step is needed to limit the precision of parameters for practical coding systems. Moreover, the normalization of GGLs makes the analysis of models with different number of vertices (dimensions) easier.}. Specifically, an estimated GGL is normalized by dividing with $w_c^\ast$ as $\smash{\widehat{\lap}} = (1/w_c^\ast)\,\lap(w_c^\ast,v_c^\ast)$. So, the edge weights of the graph associated with $\smash{\widehat{\lap}}$ are equal to 1, and its non-zero vertex weight is equal to $v_c^\ast/w_c^\ast$. Note that, the normalization of GGLs does not have an impact on their corresponding GBTs, since the set of eigenvectors (i.e., $\gbt$) in (\ref{eqn:GBT}) are invariant to scaling of $\lap$. For the rounding, a normalized vertex weight ($v_c^\ast/w_c^\ast$) is rounded to the nearest integer multiple of $0.25$. We denote the resulting rounded vertex weight as $\alpha_N$, where $N$ is the transform length (or number of vertices).

The proposed two-step approach is applied to design GBSTs as alternatives to DST-7 and DCT-8 in VVC. Since VVC allows 4, 8, 16 and 32-point DST-7/DCT-8, GBSTs are optimized for the same transform lengths. The proposed approach results in $\alpha_4 = 2$, $\alpha_8=1$, $\alpha_{16}=0.75$ and $\alpha_{32}=0.25$ for 4, 8, 16 and 32-point GBSTs, respectively. Based on the observations in Section \ref{sec:dct_dst_interpret}, the results of proposed approach can be interpreted as follows:
\begin{itemize}[noitemsep,nolistsep]
\item Since $\alpha_4=2$ for $\mathcal{L}_1$ and $\mathcal{L}_2$ corresponds to DST-4 and DCT-4, respectively, the proposed approach practically suggests replacing DST-7 with DST-4 and DCT-8 with DCT-4 for 4-point transforms to improve coding gains.
\item For 8-point transforms, the approach suggests to keep DST-7 and DCT-8, because $\alpha_8=1$ for $\mathcal{L}_1$ and $\mathcal{L}_2$ leads to DST-7 and DCT-8, respectively.
\item As $\alpha_N=0$ would indicate using DCT-2 and $\alpha_N=1$ corresponds to DST-7/DCT-8, the 16-point GBST derived from $\alpha_{16}=0.75$ leads to a transform that is closer to DST-7/DCT-8 as compared to DCT-2, yet $\alpha_{32}=0.25$ generates a 32-point GBST that is closer to DCT-2 than DST-7/DCT-8.
\end{itemize}
Fig.~\ref{fig:basis} illustrates the basis vectors of 8-point transforms for different $\alpha_8$ values. The figure demonstrates that basis vectors become more similar as their corresponding $\alpha_8$ values are closer.

\begin{table}[!t]
\centering
\caption{Coding performance of GBSTs over VTM under all-intra configurations in terms of BD-rates (BDR). Negative percentage values indicate coding gain improvements.}
\renewcommand{\arraystretch}{1.1}
\label{table:result}
\begin{tabular}{|c|l|c|c|c|} \hline
\multicolumn{2}{|c|}{Sequences}& BDR(Y) & BDR(U) & BDR(V)  \\ \hline \hline
\multirow{3}{*}{A1} & Tango2 & -0.49\% &	0.04\% & 0.05\% \\ \cline{2-5}
&FoodMarket4 & -0.58\% &	0.27\% &	0.33\%  \\ \cline{2-5}
&Campfire & -0.21\% &	0.17\% &	0.23\%  \\ \hline \hline
\multirow{3}{*}{A2}&CatRobot1 &-0.19\% &	0.39\% &	0.30\% \\ \cline{2-5}
&DaylightRoad2 & -0.34\% &	0.50\% &	0.07\% \\ \cline{2-5}
&ParkRunning3 &-0.24\% &	0.24\% &	0.30\%  \\ \hline \hline
\multirow{5}{*}{B}&MarketPlace & -0.14\% &	0.42\% &	0.45\%  \\ \cline{2-5}
&RitualDance & -0.33\% &	0.13\% &	0.20\%  \\ \cline{2-5}
&Cactus & -0.32\% &	-0.19\% &	0.02\%  \\ \cline{2-5}
&BasketballDrive & -0.42\% &	0.12\% &	-0.02\%  \\ \cline{2-5}
&BQTerrace & -0.28\% &	-0.21\% &	-0.05\%  \\ \hline \hline
\multirow{4}{*}{C}&BasketballDrill & -0.21\% &	-0.26\% &	-0.11\%  \\ \cline{2-5}
&BQMall & -0.43\% &	-0.27\% &	-0.13\%  \\ \cline{2-5}
&PartyScene & -0.41\% &	-0.44\% &	-0.40\%  \\ \cline{2-5} 
&RaceHorses & -0.34\% &	0.01\% &	-0.06\%  \\ \hline \hline
\multirow{3}{*}{E}&FourPeople & -0.43\% &	-0.27\% &	-0.43\%  \\ \cline{2-5}
&Johnny & -0.54\% &	-0.25\% &	-0.69\%  \\ \cline{2-5}
&KristenAndSara &-0.65\% &	-0.41\% &	-0.77\%  \\ \hline \hline 
\multicolumn{2}{|c|}{\textbf{Average}}& \textbf{-0.36\%} & \textbf{0.00\%} & \textbf{0.04\%} \\ \hline \hline
\multicolumn{2}{|c|}{Encoder Time[\%]} & \multicolumn{3}{|c|}{99\%} \\ \hline
\multicolumn{2}{|c|}{Decoder Time[\%]} & \multicolumn{3}{|c|}{100\%} \\ \hline
\end{tabular}
\end{table}

\begin{figure*}[!t]
\vspace{-0.25cm}
\centering
\subfigure[BDR(Y) for different $\alpha_4$ and fixed $\alpha_{8}\!=\!\alpha_{16}\!=\!\alpha_{32}\!=\!1$]{\includegraphics[width=0.39\textwidth]{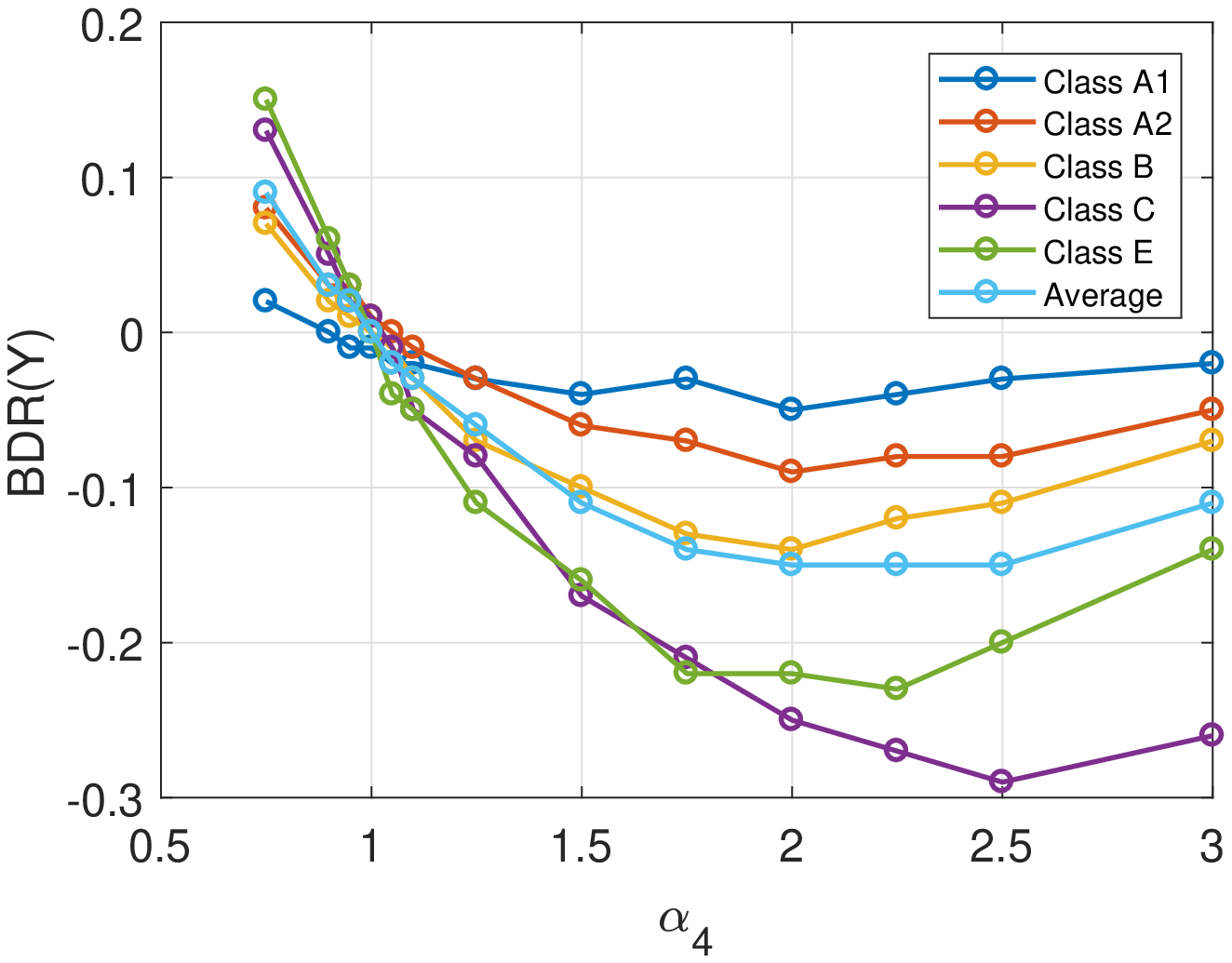}} \quad
\subfigure[BDR(Y) for different $\alpha_8$ and fixed $\alpha_{4}\!=\!\alpha_{16}\!=\!\alpha_{32}\!=\!1$]{\includegraphics[width=0.39\textwidth]{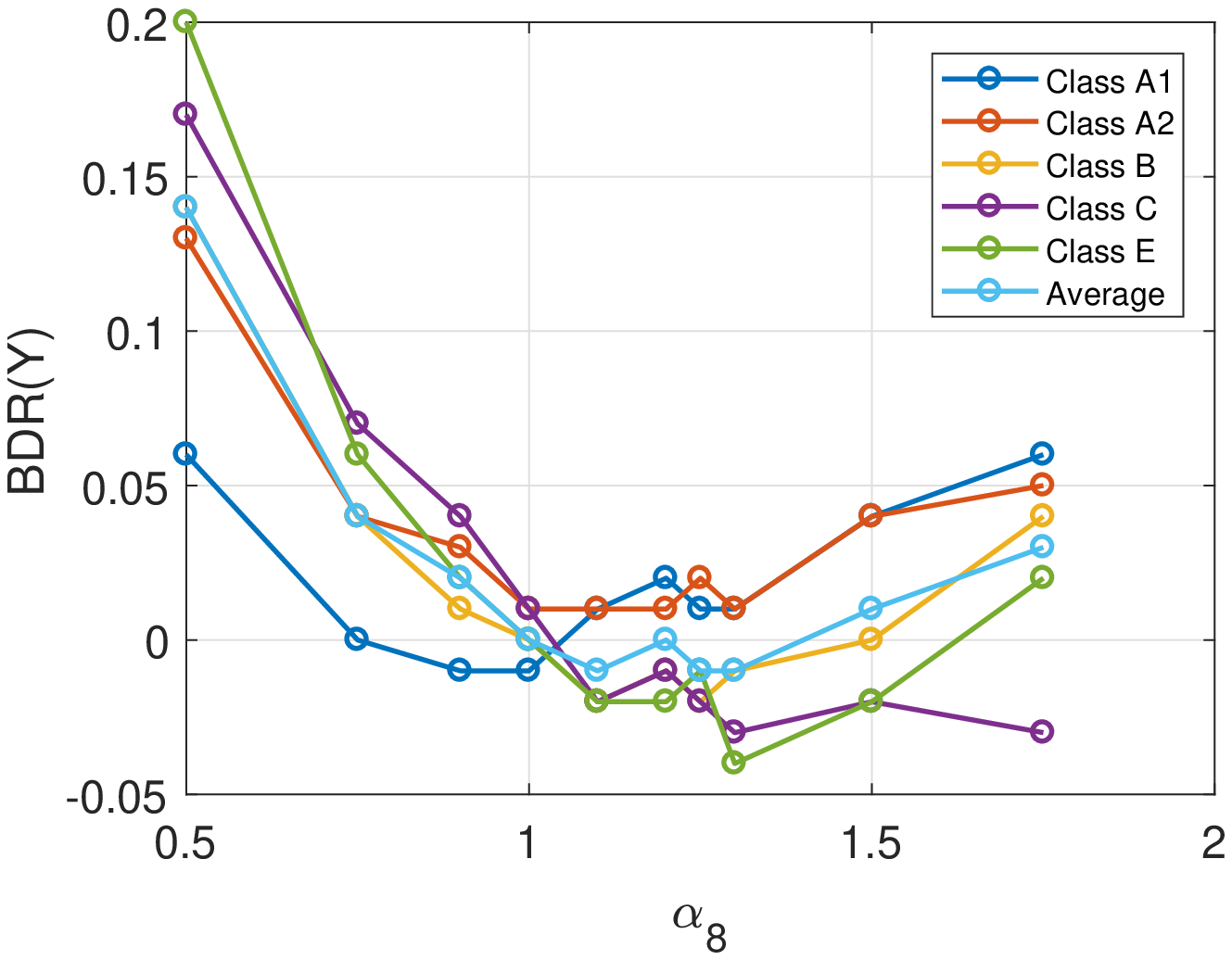}}
\subfigure[BDR(Y) for different $\alpha_{16}$ and fixed $\alpha_{4}\!=\!\alpha_{8}\!=\!\alpha_{32}\!=\!1$]{\includegraphics[width=0.39\textwidth]{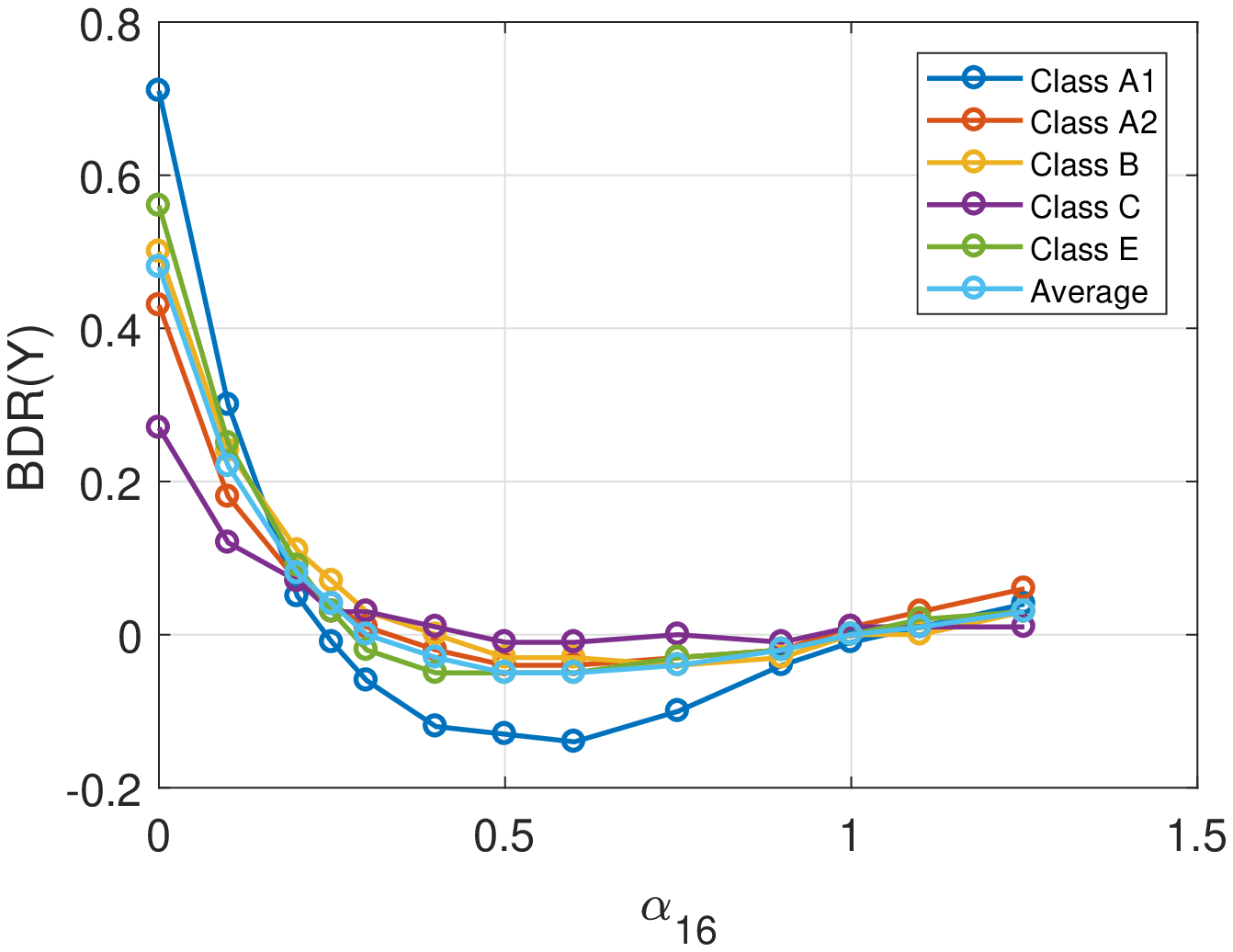}} \quad
\subfigure[BDR(Y) for different $\alpha_{32}$ and fixed $\alpha_{4}\!=\!\alpha_{8}\!=\!\alpha_{16}\!=\!1$]{\includegraphics[width=0.39\textwidth]{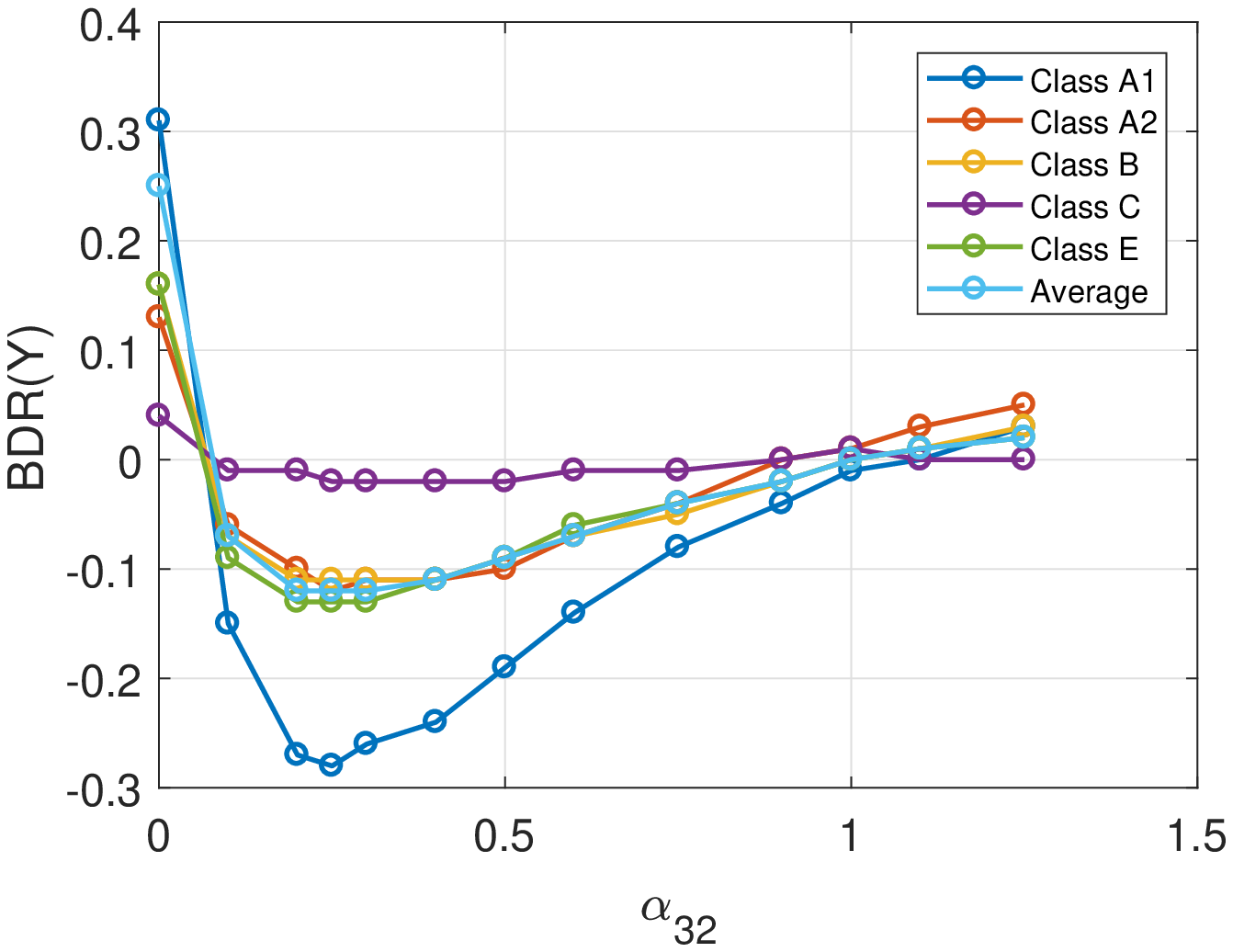}}
\caption{Average and per-class luma BD-rate performances of $N$-point GBSTs across different $\alpha_N$ values over VTM.} 
\label{fig:result}
\vspace{-0.35cm}
\end{figure*}

\section{Experimental Results}
\label{sec:results}
The performance of the proposed GBSTs are tested under common test conditions (CTC) \cite{Bossen:18:ctc} on the VVC reference software (version 2), VTM, by replacing the DST-7 and DCT-8 in MTS with the GBSTs for different transform lengths (i.e., 4, 8, 16 and 32-point transforms).  

Table \ref{table:result} shows\footnote{The results under random-access configurations (allowing inter prediction) are omitted due to limited space. Note also that MTS is disabled for inter coded blocks in CTC.} the coding gains achieved by using the proposed set of GBSTs (in Section \ref{sec:proposed_approach}) over the VTM anchor with DST-7 and DCT-8 in terms of BD-rates \cite{Bjontegaard:01:cap}. On average, the proposed method provides considerable luma coding gains of 0.36\% (in Y component). Practically, no loss is observed for chroma channel (in U and V components), since VTM only applies multiple separable transform candidates on the luma blocks. 
For all the sequences in CTC, luma coding gains are observed, and significant gains up to 0.6\% are achieved for some of the ultra-high-definition (UHD) sequences in class A1.

Fig.~\ref{fig:result} shows BD-rates of only replacing $N$-point DST-7 and DCT-8 with $N$-point GBSTs for different $\alpha_N$ values. For example, in Fig.~\ref{fig:result}c, only 16-point transforms are replaced by GBSTs derived from different $\alpha_{16}$ while keeping 4, 8 and 32-point DST-7 and DCT-8 (i.e., $\alpha_{4}=\alpha_{8}=\alpha_{32}=1$). The results empirically validate the proposed approach (described in Section \ref{sec:proposed_approach}) resulting in $\alpha_4 = 2$, $\alpha_8=1$, $\alpha_{16}=0.75$ and $\alpha_{32}=0.25$, all of which are very close to the best $\alpha_N$ with smallest BDR(Y) in Fig.~\ref{fig:result}a, \ref{fig:result}b, \ref{fig:result}c and \ref{fig:result}d, respectively. The figures also show that the impact of changing $\alpha_N$ on the BDR(Y) depends on the resolution of sequences. For example, in Fig.~\ref{fig:result}a, the coding gains for classes A1 and A2 with high-resolution are very limited, while for classes with lower resolution (e.g., class C) the gains are significant for larger $\alpha_4$. On the other hand, in Fig.~\ref{fig:result}d, smaller $\alpha_{32}$ leads to significant coding gains for classes A1 and A2, yet the gains are negligible for class C.

\section{Conclusions}
\label{sec:concl}
In this paper, we introduced a parametric approach to build graph-based separable transforms (GBSTs) better capturing the residual block statistics as compared to DST-7 and DCT-8. An optimized set of GBSTs are constructed and tested on the VVC reference software VTM by replacing the DST-7 and DCT-8 in VVC. The experimental results demonstrated that proposed GBSTs improve the coding efficiency about 0.4\% over the VTM. The proposed GBSTs do not introduce any additional hardware complexity on the current VVC design, since practical implementations of DST-7 and DCT-8 are also primarily based on matrix multiplications.

\bibliographystyle{IEEEbib}
\bibliography{ref}

\end{document}